# Questioning whether seasonal advance of intense tropical cyclones since the 1980s truly exists

Jimin Liu[1,2#], Jeremy Cheuk-Hin Leung[1,3#], Wenshou Tian[2*], Hong Huang[1], Daosheng Xu[3], Weijing Li[4], Weihong Qian[5], Banglin Zhang[1,2*]

**Affiliations**

[1] College of Meteorology and Oceanography, National University of Defense Technology, Changsha, China

[2] College of Atmospheric Sciences, Lanzhou University, Lanzhou, China

[3] Guangzhou Institute of Tropical and Marine Meteorology / Guangdong Provincial Key Laboratory of Regional Numerical Weather Prediction, CMA, Guangzhou, China

[4] National Climate Center, CMA, Beijing, China

[5] Department of Atmospheric and Oceanic Sciences, Peking University, Beijing, China

# These authors contributed equally to this work.

* Correspondence to: zhangbanglin24@nudt.edu.cn (Banglin Zhang), wstian@lzu.edu.cn (Wenshou Tian)

Shan et al.[1,2] recently reported significant seasonal advances of intense tropical cyclones (TCs) in both the Northern Hemisphere (NH) and Southern Hemisphere (SH) since the 1980s, and emphasized the data insensitivity of this conclusion, based on the Advanced Dvorak Technique–Hurricane Satellite (ADT-HURSAT)[3] and the International Best Track Archive for Climate Stewardship (IBTrACS)[4] datasets. However, this conclusion contradicts our recent findings. Following the procedures outlined in Shan et al.[1,2], our analysis reveals that the seasonal advancing trend of intense TCs does not pass the significance test in the SH. Meanwhile, for the NH, the trend is statistically significant only when using the ADT-HURSAT, but not when using the IBTrACS. These discrepancies may be due to flaws in the calculations performed by Shan et al.[1]. The above findings raise doubts about the reproducibility and validity of Shan et al.'s conclusions regarding the global seasonal advance of intense TCs. We argue that the reported seasonal advance of intense TCs since the 1980s is inconclusive, and further investigations are needed.

In Shan et al.[1,2], the authors first claimed that the trends in the median occurrence dates of intense TCs (hereafter referred to as $\widetilde{date}_{intense_TC}$) in the NH and SH are statistically significant at the 95% confidence level according to the Mann-Kendall test, based on trend analysis on the ADT-HURSAT from 1981–2017. Then, they claim that this earlier-shifting trend is independent of the choice of dataset, based on trend analysis of the IBTrACS from 1981–2020 (see Table 1 in Shan et al.[1,2]). However, we arrive at different conclusions regarding the seasonal advance of intense TCs, even if we strictly follow the definitions and analysis procedures outlined in Shan et al.[1] (see Methods).

Firstly, using the ADT-HURSAT, we obtain a statistically significant decreasing trend of $\widetilde{date}_{intense_TC}$ (-3.7±3.5 days/decade) in the NH from 1981–2017, which is consistent with the findings of Shan et al.[1]. Conversely, the $\widetilde{date}_{intense_TC}$ trend (-3.0±7.7 days/decade) in the SH does not pass the Mann-Kendall test (p=0.31, Table 1), which is inconsistent with Shan et al.[1]. It is important to note that the 95% confidence interval (7.7 days/decade, based on Student's t-test) is much larger than the trend magnitude (-3.0 days/decade), which aligns with the Mann-Kendall test result. Thus, both the Mann-Kendall test and Student's t-test yield the same conclusion that the SH $\widetilde{date}_{intense_TC}$ trend is statistically insignificant from 1981–2017.

Next, based on the IBTrACS, we find that the $\widetilde{date}_{intense_TC}$ trends are statistically insignificant in both hemispheres from 1981–2020. Note that Shan et al.[1,2] employed a 115 kt threshold, instead of 110 kt, in the trend analyses of IBTrACS (see Methods). By employing the same definition, we find that the $\widetilde{date}_{intense_TC}$ in the NH exhibits an insignificant trend (-2.4±4.5 days/decade, p=0.29) from 1981–2020. Similarly, the SH $\widetilde{date}_{intense_TC}$ trend does not pass the significance test either (-3.8±4.5 days/decade, p=0.36). These results, based on both the Mann-Kendall test and Student's *t*-test, do not support the conclusion reached by Shan et al.[1,2].

The above results show that we can replicate the $\widetilde{date}_{intense_TC}$ trend magnitudes reported by Shan et al.[1,2], with discrepancies occurring only in statistical significance (Table 1). We also notice that, despite using the same code, we successfully obtained the same 95% confidence interval for the trend analysis on ADT-HURSAT, but not for the IBTrACS. Regarding these reproducibility issues, we have communicated with Shan et al., with the help of the Editor, to ensure consistency between our codes. However, during the communication, the authors did not provide the codes used for significance test calculations. Thus, we are not allowed to tell the source of the discrepancies mentioned above. Given that our results are supported by both the Mann-Kendall test and Student's *t*-test (Table 1), we suspect that there may be some unsolved issues in their significance test calculations.

Further, the different choices of thresholds employed for identifying intense TCs across different datasets raise serious concerns. In their initial publication, Shan et al.[1,2] claimed that a 110 kt threshold was used to identify intense TCs from both the ADT-HURSAT and IBTrACS. However, upon examining Shan et al.'s code, we discovered that a 115 kt threshold was used for the IBTrACS analysis instead. After we raised this issue, the authors did not revert the threshold to 110, where the trends would have been 0.24±3.36 days/decade for the NH and -1.70±7.19 days/decade for the SH, in their Author Correction[2] (Table1). Instead, they added an unsupported statement that "It is observed that the number and proportion of intense TC cases tend to be overestimated in the best-track dataset when using the threshold value of 110 kt for intense TCs, compared to the ADT-HURSAT dataset." However, we found that this statement does not hold true. During 1981–2017, using a 110 kt threshold, we identified 398 and 153 intense TCs identified from the ADT-HURSAT dataset in the NH and SH, respectively, which are very close to the numbers identified from the IBTrACS (388 for NH, 144 for SH). In contrast, the number of intense TCs identified from the IBTrACS using a 115 kt threshold is much smaller (316 for NH, 110 for SH). It appears that a 115 kt threshold was chosen for the IBTrACS analysis because it yields a larger $\widetilde{date}_{intense_TC}$ trend values compared to a 110 kt threshold (Table 1).

Nonetheless, no matter whether a 110 kt or 115 kt threshold is employed, our results reveal that the conclusion "the $\widetilde{date}_{intense_TC}$ trends are statistically significant in both hemispheres" is not valid, contradicting Shan et al.'s conclusion. If using a 110 kt, as discussed above, a statistically significant $\widetilde{date}_{intense_TC}$ trend (-3.7±3.5 days/decade) is observed only in the NH, but not in the SH, based on the ADT-HURSAT. The $\widetilde{date}_{intense_TC}$ trends based on the IBTrACS are insignificant in both hemispheres. Similarly, if using a 115 kt, the $\widetilde{date}_{intense_TC}$ trend (-5.2±4.6 days/decade) is statistically significant only in the NH, based on the ADT-HURSAT (Table 1).

To sum up, this commentary raises a reproducibility issue about the conclusions reported in Shan et al.[1,2]. While Shan et al.[1,2] concluded that the $\widetilde{date}_{intense_TC}$ trends in both hemispheres are significant, based on the ADT-HURSAT and IBTrACS, we find that this conclusion is not replicable. This discrepancy may be due to flaws in the significance test calculation performed by Shan et al. This reproducibility issue is a crucial concern raising substantial doubts about the existence of seasonal advance of intense TCs.

In studying the long-term variability of TC activity, the ADT-HURSAT is often considered a more reliable data source due to its temporal homogeneity, compared to the IBTrACS. However, our results suggest that even using the ADT-HURSAT, the seasonal advancing trend of intense TCs is significant only in the NH, but not in both hemispheres as claimed in Shan et al.[1]. Therefore, we argue that there is insufficient evidence to support the conclusion of intense TC seasonal advance. The reported global seasonal advance of intense TCs since the 1980s is inconclusive and we strongly suggest the authors clarify this issue.

## Tables

**Table 1 | Reproducibility and data sensitivity of seasonal advance in intense TCs.** Linear trends (unit: days / decade), and the corresponding 95% confidence interval, of $\widetilde{date}_{intense_TC}$ in the NH and SH, based on the IBTrACS and ADT-HURSAT, respectively. Asterisks indicate the confidence levels (1 asterisk = 90%, 2 asterisks = 95%, and 3 asterisks = 99%). The p-values of linear trends are calculated based on the Student's t-test and Mann-Kendall test. The 95% confidence intervals of linear trends were estimated by the Student's *t*-test. Blue fonts denote results that are inconsistent between our analyses and the results of Shan et al.[1,2]. Results show that the $\widetilde{date}_{intense_TC}$ trend is statistically significant only in the NH when using the ADT-HURSAT (bold text).

| Dataset (LMI / Period) | Hemisphere | Intense TC count | Original results in Shan et al.[1] (p-values based MK test) | Updated results in Shan et al.[2] (p-values based MK test) | Results in this study (p-values based on t-test/MK test) |
|---|---|---|---|---|---|
| ADT-HURSAT (LMI > 110 / 1981–2017) | NH | 398 | **-3.7**±3.6 (< 0.05)** | **-3.7**±3.5 (< 0.05)** | **-3.70**±3.52 (0.04/0.03)** |
| | SH | 153 | -3.2**±7.9 (< 0.05) | -3.0**±7.7 (< 0.05) | -3.00±7.66 (0.43/0.31) |
| ADT-HURSAT (LMI > 115 / 1981–2017) | NH | 328 | Not given | Not given | -5.22**±4.61 (0.03/0.02) |
| | SH | 120 | Not given | Not given | -0.47±9.15 (0.92/0.94) |
| IBTrACS (LMI > 110 / 1981–2017) | NH | 388 | Not given | Not given | -0.46±3.87 (0.81/0.55) |
| | SH | 144 | Not given | Not given | -2.12±8.40 (0.61/0.82) |
| IBTrACS (LMI > 115 / 1981–2017) | NH | 316 | Not given | Not given | -3.93±5.07 (0.13/0.10) |
| | SH | 110 | Not given | Not given | -5.07±9.73 (0.30/0.31) |
| IBTrACS (LMI > 110 / 1981–2020) | NH | 431 | -2.3**±2.1 (< 0.05) | Not given | 0.24±3.36 (0.89/0.94) |
| | SH | 158 | -4.1**±3.9 (< 0.05) | Not given | -1.70±7.19 (0.63/0.94) |
| IBTrACS (LMI > 115 / 1981–2020) | NH | 350 | Not given | -2.4**±2.2 (< 0.05) | -2.39±4.48 (0.29/0.29) |
| | SH | 120 | Not given | -3.8**±4.2 (< 0.05) | -3.83±8.33 (0.36/0.45) |

## Methods

*Data*

The TC datasets, namely the ADT-HURSAT and IBTrACS datasets, used in this commentary are the same as those employed in Shan et al.[1]. The ADT-HURSAT dataset offers satellite-derived TC information from 1981–2017, and can be obtained from the Supporting Information section of Kossin et al.[3]. The IBTrACS provides official TC track records from 1981–2020 for this study, and it is accessible through https://www.ncei.noaa.gov/products/international-best-track-archive.

*Identification of intense TCs*

Following Shan et al.[1], a TC is considered intense when its lifetime maximum wind speed is greater than 110 kt. However, since Shan et al.[1,2] applied a different threshold (115 kt) for the analyses on IBTrACS in their Author Correction, we also discuss the results based on the 115 kt threshold. Only intense TCs in the North Atlantic (NA), East Pacific (EP), West Pacific (WP) and South Indian (SI), and South Pacific (SP) were analyzed [1].

*Active TC season*

While it would be more reasonable to define active TC season of each basin based on the TC occurrence climatology[5,6], we strictly follow the definition given in Shan et al.[1] in this study for the convenience of comparison. Namely, the NH TC season lasts from 1 January to 31 December. The SH TC season lasts from 1 July of the last calendar year to 30 June of the current calendar year. The active TC season of NH lasts from 1 June to 30 November, and that of SH lasts from 1 December of the last calendar year to 30 April of the current calendar year.

*Median of intense TC occurrence dates ($\widetilde{date}_{intense_TC}$)*

The median value of intense TC occurrence dates ($\widetilde{date}_{intense_TC}$) of each active TC season is estimated for each hemisphere and each basin. Specifically, we first determine the occurrence date of each detected intense TC, which is defined as the first time when the TC reaches its lifetime maximum intensity (LMI). Then, for each active TC season, the $\widetilde{date}_{intense_TC}$ is estimated as the 50th percentile value of the intense TC occurrence dates of that season.

*Trend analysis and significance test*

All trend values were calculated using linear regression. The statistical significance of all linear trend analyses was tested by the non-parametric Mann-Kendall test and the Student's t-test. The 95% confidence intervals of linear trends were estimated by the Student's *t*-test.

## Data availability

The ADT-HURSAT dataset can be obtained from the Supporting Information section of Kossin et al.[3], and the IBTrACS dataset (v4.01) is accessible through https://www.ncei.noaa.gov/products/international-best-track-archive.

## Code availability

All the codes employed in this uploaded to Code Ocean during the submission.

## Acknowledgments

This study was supported by the NUDT Research Initiation Funding for High-Level Scientific and Technological Innovative Talents (202402-YJRC-LJ-001), the National Natural Science Foundation of China (42405038), and the Guangdong Province Introduction of Innovative R&D Team Project China (2019ZT08G669).

## Competing interests

The authors declare no competing interests.

## Author contributions

J.L: methodology, formal analysis, writing—review and editing, visualization; J.C.H.L.: methodology, formal analysis, writing—original draft, writing—review and editing; W.T., supervision, writing—review and editing, funding acquisition; H.H., D.X., W.Q., W.L.: writing—review and editing; B.Z.: conceptualization, supervision, methodology, writing—original draft, writing—review and editing, funding acquisition.